\documentstyle[12pt]{article} 
\makeatletter \def\@cite#1#2{{[{#1}]\if@tempswa\typeout {IJCGA
warning: optional citation argument ignored: `#2'} \fi}}

\newcount\@tempcntc
\def\@citex[#1]#2{\if@filesw\immediate\write\@auxout{\string\citation{#2}}\fi
  \@tempcnta\z@\@tempcntb\m@ne\def\@citea{}\@cite{\@for\@citeb:=#2\do
    {\@ifundefined
       {b@\@citeb}{\@citeo\@tempcntb\m@ne\@citea\def\@citea{,}{\bf ?}\@warning
       {Citation `\@citeb' on page \thepage \space undefined}}%
    {\setbox\z@\hbox{\global\@tempcntc0\csname b@\@citeb\endcsname\relax}%
     \ifnum\@tempcntc=\z@ \@citeo\@tempcntb\m@ne
       \@citea\def\@citea{,}\hbox{\csname b@\@citeb\endcsname}%
     \else
      \advance\@tempcntb\@ne
      \ifnum\@tempcntb=\@tempcntc
      \else\advance\@tempcntb\m@ne\@citeo
      \@tempcnta\@tempcntc\@tempcntb\@tempcntc\fi\fi}}\@citeo}{#1}}
\def\@citeo{\ifnum\@tempcnta>\@tempcntb\else\@citea\def\@citea{,}%
  \ifnum\@tempcnta=\@tempcntb\the\@tempcnta\else
   {\advance\@tempcnta\@ne\ifnum\@tempcnta=\@tempcntb \else 
\def\@citea{--}\fi
    \advance\@tempcnta\m@ne\the\@tempcnta\@citea\the\@tempcntb}\fi\fi}
\makeatother

\def\boxit#1{\leavevmode\thinspace\hbox{\vrule\vtop{\vbox{\hrule%
        \vskip3pt\kern1pt\hbox{\vphantom{\bf/}\thinspace\thinspace%
        {\bf#1}\thinspace\thinspace}}\kern1pt\vskip3pt\hrule}\vrule}%
        \thinspace}
\def\Boxit#1{\noindent\vbox{\hrule\hbox{\vrule\kern3pt\vbox{
\advance\hsize-7pt\vskip-\parskip\kern3pt\bf#1 \hbox{\vrule height0pt
depth\dp\strutbox width0pt} \kern3pt}\kern3pt\vrule}\hrule}}

\newcommand{\Hh}{\lower1.2ex\hbox{$\stackrel{\textstyle
H}{\footnotesize\sim}$}}
\newcommand{\Hho}{\lower1.2ex\hbox{$\stackrel{\textstyle
H_1}{\footnotesize\sim}$}}
\newcommand{\Hhw}{\lower1.2ex\hbox{$\stackrel{\textstyle
H_2}{\footnotesize\sim}$}}
\newcommand{\h}{\lower1.2ex\hbox{$\stackrel{\textstyle
h}{\footnotesize\sim}$}}

\newcommand{\gsim}{\lower.7ex\hbox{$\;\stackrel{\textstyle>}{\sim}\;$}}
\newcommand{\lsim}{\lower.7ex\hbox{$\;\stackrel{\textstyle<}{\sim}\;$}}
\newcommand{\be}{\begin{equation}} \newcommand{\ee}{\end{equation}}
\newcommand{\beq}{\begin{equation}} \newcommand{\eeq}{\end{equation}}
\newcommand{\bea}{\begin{eqnarray}} \newcommand{\eea}{\end{eqnarray}}

 \def\bma#1{\mbox{\boldmath{$#1$}}}
\def\simlt{\stackrel{<}{{}_\sim}}

\def\baselinestretch{1}
\begin{document}
\catcode`@=11 \newtoks\@stequation
\def\subequations{\refstepcounter{equation}%
\edef\@savedequation{\the\c@equation}%
\@stequation=\expandafter{\theequation}
\edef\@savedtheequation{\the\@stequation}
\edef\oldtheequation{\theequation}
\def\theequation{\oldtheequation\alph{equation}}}
\def\endsubequations{\setcounter{equation}{\@savedequation}%
\@stequation=\expandafter{\@savedtheequation}%
\edef\theequation{\the\@stequation}\global\@ignoretrue

\noindent} \catcode`@=12
\begin{titlepage}

\title{{\bf  
Scalar Loops in Little Higgs Models}} 
\vskip3in \author{ {\bf\sc J.R. Espinosa} and
{\bf\sc J.M. No\footnote{E-mail 
addresses: {\tt jose.espinosa@uam.es, josemi.no@uam.es}}}
\hspace{3cm}\\
{\small IFT-UAM/CSIC, 28049 Madrid, Spain}.
}  \date{}  \maketitle  \def\baselinestretch{1.15}
\begin{abstract}
\noindent
Loops of the scalar particles present in Little Higgs models generate 
radiatively scalar operators that have been overlooked before in Little 
Higgs analyses. We compute them using a technique, recently proposed 
to deal with scalar fluctuations in non-linear sigma models, that greatly 
simplifies the calculation. In particular models some of these operators 
are not induced by loops of gauge bosons or fermions, are consistent 
with the Little Higgs symmetries that protect the Higgs boson mass, and
must also be included in the Lagrangian. In general, scalar loops 
multiplicatively renormalize the tree-level scalar operators, ${\cal 
O}_S\rightarrow {\cal O}_S [1- {\cal N} \Lambda^2/(4\pi f)^2]$ with large 
${\cal N}$ ({\it e.g.}~${\cal N}\sim 20$ for the Littlest Higgs), 
suggesting a true UV cutoff $\Lambda \simlt 4\pi f 
/\sqrt{{\cal N}}$ significantly below the estimate $4\pi f$ of naive dimensional analysis.
This can have important 
implications for the phenomenology and viability of Little Higgs models.
\end{abstract}

\thispagestyle{empty}
\vspace*{4cm} \leftline{October 2006} \leftline{}

\vskip-19cm \rightline{IFT-UAM/CSIC-06-49} 
\rightline{hep-ph/0610255} \vskip3in

\end{titlepage}
\setcounter{footnote}{0} \setcounter{page}{1}
\newpage
\baselineskip=20pt

\noindent

\section{Introduction}

The Little Hierarchy problem concerns the tension between the electroweak 
scale (determined by the Higgs mass parameter in the Lagrangian) and the 
10 TeV scale, which is roughly the minimal suppression scale of 
non-renormalizable 
operators required not to upset the fit to precision electroweak data. The 
little hierarchy between these two scales would require a fine-tuning of 
${\cal O}(1\%)$ if the low-energy effective theory below $\Lambda\sim 10$ 
TeV is the pure Standard Model (SM).

Little Higgs (LH) models 
\cite{Littlest,Simplest,SU6Sp6,ChengLow,HS,Schmaltz,Wacker} 
try to solve 
this little hierarchy problem by 
making the Higgs doublet a pseudo-Goldstone of some global symmetry $G$ 
which is broken (both explicitly and spontaneously) in a smart way 
(``collective breaking'') that keeps the Higgs mass protected from 1-loop 
quadratically divergent corrections. Implementing such global symmetry 
requires new fields beyond those of the SM: new gauge bosons, fermions and 
scalars, filling out multiplets of $G$. The spontaneous breaking 
has an order parameter $f\sim 1$ TeV and the new LH particles gain masses 
of that order. In most existing LH models one treats this breaking in an 
effective way using a non-linear sigma model description so that one keeps 
in the low-energy effective theory the (pseudo-)Goldstone scalars, which 
remain light after the spontaneous breaking of $G$. The dynamics of this 
breaking belongs in the complete theory at $\Lambda\sim 4\pi f\sim 10$ 
TeV, which acts as the limit of validity of the low-energy effective 
theory.

The Higgs doublet is among these pseudo-Goldstone fields but, due to the 
particular way in which the breaking is arranged, it does not get a mass 
of order $f\sim 1$ TeV but has a special protection and its mass squared 
is of order $f^2/(16\pi^2)$ which, at least parametrically, is of the 
order of the electroweak scale, as desired. The LH scalar sector is 
usually treated in the following way. Calling $\Sigma$ the scalar 
matrix that contains the pseudo-Goldstone degrees of freedom (in some kind 
of non-linear exponential parametrization of fluctuations around the vacuum 
$\Sigma_0$) the low-energy Lagrangian for $\Sigma $ is basically 
determined by symmetry considerations and can be organized as a momentum 
expansion. The kinetic part is of the form
\be
{\cal L}_k = \frac{f^2}{8} 
{\mathrm Tr}[(D_{\mu}\Sigma)(D^{\mu}\Sigma)^\dagger]\ ,
\label{Lk}
\ee
where $D_{\mu}\Sigma$ is the covariant derivative, which depends on the 
gauged subgroups of $G$ in any given model.

The matrix $\Sigma$ also couples to fermions respecting the symmetries 
that protect the Higgs mass. In a generic $\Sigma$-background, fermions 
and 
gauge bosons will have $\Sigma$-dependent mass matrices ${\cal 
M}_F(\Sigma)$ and ${\cal M}_V(\Sigma)$. Loops of these particles will then 
generate scalar operators. In fact, the one-loop scalar potential contains 
quadratically divergent contributions of the form
\be
\delta_q V={\Lambda^2\over 32\pi^2}{\mathrm Str}{\cal M}^2\ ,
\label{dqV}
\ee
where ${\mathrm Str}$ traces over degrees of freedom with negative sign 
for fermions. Plugging ${\cal M}_F(\Sigma)$ and ${\cal M}_V(\Sigma)$ in 
(\ref{dqV}) and using the estimate of naive dimensional analysis (NDA) 
$\Lambda\sim 4\pi f$ \cite{NDA}, one gets two types of 
$\Sigma$-operators:
\bea
{\cal O}_V(\Sigma)& \propto & f^2 \ {\mathrm Tr}\left[ {\cal 
M}_V^2(\Sigma)\right]
\ ,\nonumber\\
{\cal O}_F(\Sigma)& \propto & f^2 
\ {\mathrm Tr} \left[{\cal M}_F^\dagger(\Sigma) {\cal M}_F(\Sigma)\right]\ 
,
\eea
which one should include in the Lagrangian from the very beginning as they 
would be generated radiatively anyway. So one writes
\be
\delta_S{\cal L}=c_V\ {\cal O}_V(\Sigma)+c_F\  {\cal O}_F(\Sigma)\ ,
\label{LS}
\ee 
with some unknown coefficients $c_V$ and $c_F$ expected to be ${\cal 
O}(1)$ (although in particular models these might turn out to be sizeable 
\cite{CEH}).

In fact such operators are crucial for the viability of LH models: they 
generate ${\cal O}(f)$ masses for all pseudo-Goldstones other than the 
doublet Higgs, and scalar quartic couplings, including that of the Higgs doublet. 
In the presence of the non-derivative scalar interactions of (\ref{LS}) 
one might wonder about the scalar contribution to (\ref{dqV}), which is 
missing in LH analyses in the literature. Analyzing this issue is the 
main motivation of this paper.

Perhaps one of the reasons for overlooking this scalar contribution is that its 
calculation is not as easy as that for the scalar operators induced by 
fermion and gauge boson loops. To compute the contribution from scalar 
loops one would need first to find ${\cal M}_S^2(\Sigma)$, the squared 
mass matrix for scalar fluctuations in a general $\Sigma$-background. One 
immediate complication is that taking derivatives of (\ref{LS}) with 
respect to scalar fields [to get ${\cal M}_S^2(\Sigma)$] destroys the 
$\Sigma$ structure so that it seems impossible to write down the matrix 
elements ${\cal M}_S^2(\Sigma)_{ab}$ as functions of $\Sigma$. Somehow one 
expects that all the bits and pieces in ${\cal M}_S^2(\Sigma)_{ab}$ will 
arrange themselves so as to give a ${\mathrm Tr}\ [{\cal M}_S^2(\Sigma)]$ 
that can indeed be expressed as a function of $\Sigma$.

To complicate things further, in a generic $\Sigma$-background the 
kinetic terms in (\ref{Lk}) are not canonical. The $\Sigma$-dependent 
re-scaling of scalar fluctuations required to get back to canonical 
kinetic terms also affects the form of ${\cal M}_S^2(\Sigma)$ entering in 
(\ref{dqV}). The programme then might be straightforward but seems rather 
cumbersome and tedious. Fortunately, a method recently developed in 
\cite{ELR} to deal efficiently with scalar fluctuations in non-linear 
sigma models, is ideally suited for our task. 

We explain this method in 
the next section, using as an illustrative example the Littlest 
Higgs model \cite{Littlest}, and calculate ${\mathrm Tr}\ [{\cal 
M}_S^2(\Sigma)]$ to see what scalar 
operators are generated in this way. In section~3 we confront the 
scalar-induced $\Sigma$-operators with the operators generated by loops of 
gauge bosons and fermions discussing their symmetry properties and 
their expansions in terms of physical fields. In section~4 we examine some 
of the implications of these scalar loop corrections to the effective 
potential.  In particular, we show how the appearance of 
scalar-induced operators can be used to set an upper bound on the scale 
$\Lambda$ up to which the LH effective theory is valid. From purely 
low-energy arguments, it follows that such cutoff is in general 
significantly lower 
than $4\pi f$  ({\it e.g.}, it is $\sim 4\pi f/\sqrt{20}$ for the Littlest 
Higgs model). This can have implications for the ability of LH models to 
solve the little hierarchy problem. Some implications for vacuum 
alignment issues are also discussed. In section~5 we present  some conclusions.
Appendix~A is devoted to logarithmically divergent corrections from scalar 
loops using again as example the Littlest Higgs model. Appendix~B deals 
with scalar loop corrections in a different LH model, based on 
$SU(6)/Sp(6)$, as an example in which scalar-induced operators are truly 
new and different from those generated by loops of gauge bosons or 
fermions. Finally, Appendix~C examines a toy $SU(N)/SO(N)$ LH model to 
investigate how the new bound on the UV cut-off derived in this paper
scales with the size of the groups involved.
 
\section{An Efficient Way of Dealing with Scalar Fluctuations}

It is convenient to use a concrete model to illustrate the method we 
follow to compute in an efficient way ${\mathrm Tr}\ [{\cal 
M}_S^2(\Sigma)]$. We use for that purpose the original Littlest Higgs 
model \cite{Littlest,Littlestmore} even if it is difficult to reconcile 
with electroweak precision tests (such issues are not relevant for 
the method itself and we prefer to keep the model simple). The model is 
based 
on a non-linear sigma model with coset structure $SU(5)/SO(5)$.  The 
spontaneous 
breaking of $SU(5)$ down to $SO(5)$ is produced by the vacuum expectation 
value (VEV) of a $5\times 5$ symmetric matrix $\Psi$ [which transforms 
under $SU(5)$ as $\Psi\rightarrow U\Psi U^T$], {\it e.g.}~when 
$\langle\Psi\rangle=I_5$ (we call $I_n$ the $n\times n$ identity 
matrix). The breaking of this global $SU(5)$ symmetry produces 14 
Goldstone 
bosons among which lives the scalar Higgs field. Following 
\cite{Littlest}, 
we make a basis change from $\langle\Psi\rangle=I_5$ to 
$\langle\Psi\rangle=\Sigma_0$ where
\beq
\Sigma_0 =\left(\begin{array}{ccc}
0 &0&I_2\\
0&1&0\\
I_2&0&0
\end{array}\right)\ .
\label{vev}
\eeq
Let us call $U_0$ the $SU(5)$ matrix that performs this change of basis. 
One has then $\Sigma_0=U_0U_0^T$ and all the group generators transform as 
$t_a=U_0t_a^{(0)}U_0^\dagger$ [$t_a^{(0)}$ are the generators in the 
original basis]. The unbroken $SO(5)$ generators satisfied the obvious 
relation $T_a^{(0)}+T_a^{(0)T}=0$ and multiplying on the left by $U_0$ and 
on the right by $U_0^T$ we arrive at the condition
\be
T_a\Sigma_0+\Sigma_0T_a^T=0\ ,
\ee
for the generators in the new basis (a condition which is immediate to
obtain alternatively just by requiring invariance of $\Sigma_0$). In the
original basis the broken generators obviously satisfy
$X_a^{(0)}=X_a^{(0)T}$.  Multiplying again by $U_0$ and $U_0^T$ one gets 
in the transformed basis
\be
X_a\Sigma_0=\Sigma_0 X_a^T\ .
\ee
The Goldstone boson degrees of freedom can be parametrized through the 
nonlinear sigma model field
\beq
\Sigma \equiv e^{i \Pi/f} \Sigma_0 e^{i \Pi^T/f} = e ^{2i\Pi/f}\Sigma_0,
\label{Sigma}
\eeq
where $\Pi = \sum_a \pi_a X_a$ and we can choose the $\pi_a$ as real 
fields (in which case $a$ runs from 1 to 14). The model has a gauged 
$SU(2)_1\times SU(2)_2\times U(1)_1 \times U(1)_2$ subgroup of $SU(5)$ 
with generators
\beq
Q_1^\alpha = \left(\begin{array}{cc}
\sigma^\alpha/2 & \\
 & 0_3
\end{array}\right)\ , \hspace{0.5cm}  
Q_2^\alpha = \left(\begin{array}{cc}
0_3 & \\
 & -\sigma^{\alpha^{*}}/2
\end{array}\right)\ ,
\eeq
(where $\sigma^\alpha$ are the Pauli matrices) and 
\beq
Y_1={\mathrm diag}(-3,-3,2,2,2)/10\ , \hspace{0.5cm}
Y_2={\mathrm diag}(-2,-2,-2,3,3)/10\ .
\eeq 
The VEV in 
eq.~(\ref{vev})
additionally breaks $SU(2)_1\times SU(2)_2\times U(1)_1 \times U(1)_2$ 
down to the SM $SU(2)_L\times U(1)_Y$ group.  

The hermitian matrix $\Pi$ in eq.~(\ref{Sigma}) contains the Goldstone and 
(pseudo)-Goldstone bosons:
\beq
\Pi = \left(\begin{array}{ccc}
\xi &\frac{H^{\dagger}}{\sqrt 2}&\Phi^{\dagger}\\
\frac{H}{\sqrt 2}&0&\frac{H^{*}}{\sqrt 2}\\
\Phi&\frac{H^{T}}{\sqrt 2}&\xi^T
\end{array}\right)+{1\over \sqrt{20}}\zeta^0{\mathrm diag}(1,1,-4,1,1)\ ,
\label{pi}
\eeq
where $H=(h^0,h^+)$ is the Higgs doublet; $\Phi$ is a complex $SU(2)$ 
triplet given by the symmetric $2\times 2$ matrix:
\be
\Phi=\left[
\begin{array}{cc}
\phi^0 & {1\over \sqrt{2}} \phi^+\\
{1\over \sqrt{2}} \phi^+ & \phi^{++}
\end{array}
\right]\ ,
\ee
the field $\zeta^0$ is a singlet which is the Goldstone associated to 
$U(1)_1 \times U(1)_2\rightarrow U(1)_Y$ breaking and finally, 
$\xi$ is the real triplet of 
Goldstone bosons associated to $SU(2)_1\times SU(2)_2\rightarrow SU(2)$ 
breaking:
\be
\xi={1\over 2} \sigma^\alpha\xi^\alpha=\left[
\begin{array}{cc}
{1\over 2}\xi^0 & {1\over \sqrt{2}} \xi^+\\
{1\over \sqrt{2}} \xi^- & -{1\over 2}\xi^{0}
\end{array}
\right]\ .
\ee

The kinetic part of the Lagrangian  is of the general form (\ref{Lk}) with
\beq
D_{\mu}\Sigma = \partial_{\mu} \Sigma 
- i\sum_{j=1}^2 g_j W_{j\mu}^\alpha(Q_j^\alpha \Sigma +
\Sigma Q_j^{\alpha T}) 
 - i\sum_{j=1}^2 g_j' B_{j\mu}(Y_j \Sigma + \Sigma Y_j^T).
\eeq
In this model, additional fermions  are introduced as a vector-like coloured
pair $\tilde{t}, \tilde{t}^c$ to cancel the quadratic divergence from top
loops (we neglect the other small Yukawa couplings). The relevant part of the
Lagrangian containing the top Yukawa coupling is given by
\beq
{\cal L}_f = {1\over 2}h_1 f \epsilon_{ijk} \epsilon_{xy} \chi_i 
\Sigma_{jx} 
\Sigma_{ky} 
u'^{c}_{3} + h_2 f \tilde{t} \tilde{t}^c + h.c.,
\eeq
where $\chi_i = (b, t,\tilde{t})$, indices $i,j,k$ run from 1 to 3 and 
$x,y$ from 4 to 5, and $\epsilon$ is the completely antisymmetric tensor.

As explained in the introduction, considering gauge and fermion loops, one
sees that the Lagrangian should also include gauge invariant terms of the
form,
\bea
-{\cal L}_S =  V  & = & 2 c_{V_i} f^4 g_i^2 \sum_\alpha 
{\mathrm Tr}[(Q_i^{\alpha}\Sigma)(Q_i^{\alpha}\Sigma)^*]  + 2 c_Y f^4 {g'}^2 
{\mathrm Tr}[(Y\Sigma)(Y\Sigma)^*]\nonumber\\
& +& 4 c_F f^4 h_1^2 {\mathrm Tr} 
[\Sigma_1\epsilon\Sigma_1^T\Sigma_1^*\epsilon\Sigma_1^\dagger]\ ,
\label{potential}
\eea
where $\Sigma_1$ is a $3\times 2$ matrix defined by
\be
\Sigma_1\equiv (\Sigma_{ix}) \ ,
\ee
with $i=\{1,2,3\}$, $x=\{4,5\}$
and $\epsilon$ is the $2\times 2$ completely antisymmetric tensor so that
\be
{\mathrm Tr} 
[\Sigma_1\epsilon\Sigma_1^T\Sigma_1^*\epsilon\Sigma_1^\dagger]=
\epsilon^{wx}\epsilon_{zy}\Sigma_{iw}\Sigma_{jx}
\Sigma^{iy *}\Sigma^{jz *}\ .
\ee
Finally, $c_{V_i}$, $c_Y$ and $c_F$ are assumed to be constants of ${\cal 
O}(1)$.

This Lagrangian produces a mass of order $f$ for the gauge bosons
($W'$) associated to the broken (axial) $SU(2)$, for a vector-like
combination of $\tilde{t}$ and $u_3'$ and for the complex scalar $\Phi$.  
Finally, the Higgs boson has zero tree level mass but a nonzero 
quartic coupling.

Let us now discuss the method introduced in \cite{ELR} to compute the 
masses of scalar fluctuations ({\it i.e.}~the degrees of freedom in 
$\Sigma$ itself) for instance in a simple background $h=\sqrt{2}\langle 
h^0\rangle$. The standard way of doing this would be to shift 
$h^0\rightarrow h^0+h/\sqrt{2}$ in $\Sigma$ to compute $h$-dependent 
scalar masses. As explained in the Introduction, to do this properly one 
should take into account that in general, after this shifting, the scalar 
kinetic terms from (\ref{Lk}) are not canonical. Therefore one should 
re-scale the fields to get the kinetic terms back to canonical form and 
this re-scaling affects the $h$-dependent contributions to scalar 
masses. Instead of following this standard procedure, ref.~\cite{ELR} used 
an alternative method which simplifies the calculations and has 
many appealing 
features.

The idea is to treat the new background with $\langle H\rangle 
=h/\sqrt{2}$ (or any other generic background) as a basis change (recall 
the discussion above of the change from $\langle\Phi\rangle=I_5$ to 
$\langle\Phi\rangle=\Sigma_0$). The $SU(5)$ transformation to the new 
background is now $U_b\equiv \exp (i\langle \Pi\rangle/f)$ with 
$\Sigma_b\equiv \langle \Sigma \rangle=U_b \Sigma_0 U_b^T$. To parametrize 
the scalar fluctuations around this background one again uses the 
exponentials of broken generators, but taking into account the effect of 
the change of basis, which acts on generators as $X_a\rightarrow U_b X_a 
U_b^\dagger$. That is, instead of using $\exp(i\Pi/f)$ one uses $U_b 
\exp(i\Pi/f) U_b^\dagger$, and writes for $\Sigma$:
\be
\Sigma = (U_b e^{i\Pi/f} U_b^\dagger)(U_b \Sigma_0 U_b^T)(U_b^* 
e^{i\Pi^T/f} U_b^T)
= e^{i\langle\Pi\rangle/f} e^{2i\Pi/f}e^{i\langle\Pi\rangle/f} 
\Sigma_0 \ . 
\label{good}
\ee
The prescription in eq.~(\ref{good}), which we could call 
``multiplicative'', is to be compared with the standard (``additive'') 
procedure
\be
\Sigma = e^{2i(\Pi+\langle\Pi\rangle)/f}\Sigma_0 \ .
\label{bad}
\ee
As already discussed in \cite{ELR}, it is easy to check that with the 
prescription of eq. (\ref{good}) scalar fluctuations are automatically 
canonical. Second, one is  free to choose this parametrization of scalar 
fluctuations: a general theorem \cite{coord} guarantees that this 
different parametrization does not change the physics. 

The parametrization above will be most useful for our goal in this 
paper\footnote{The same technique can be applied to the calculation of 
logarithmically divergent corrections, see Appendix~A.}. 
We need to compute scalar masses in a generic $\Sigma$-background. As 
${\mathrm Tr} [{\cal M}^2(\Sigma)]$ is linear in ${\cal M}^2(\Sigma)$ we 
can 
compute separately the contributions to the trace coming from the scalar operators 
${\cal O}_{V_i}(\Sigma)$, ${\cal O}_Y(\Sigma)$ and ${\cal O}_F(\Sigma)$ 
separately. For the gauge boson induced operators we have to compute 
$\delta_V {\mathrm Tr}[{\cal M}_S^2(\Sigma)]$, which is given by
\bea
2c_{V_i} f^4 g_i^2 \sum_a {\mathrm Tr} \left[ 
(Q_i^{\alpha}\partial_a^2\Sigma)(Q_i^{\alpha}\Sigma)^*
+2(Q_i^{\alpha}\partial_a\Sigma)(Q_i^{\alpha}\partial_a\Sigma)^*
+(Q_i^{\alpha}\Sigma)(Q_i^{\alpha}\partial_a^2\Sigma)^*\right]
\nonumber\\
 + 2 c_{Y_i} f^4 {g_i'}^2 \sum_a
{\mathrm Tr}\left[(Y_i\partial_a^2\Sigma)(Y_i\Sigma)^*
+2(Y_i\partial_a\Sigma)(Y_i\partial_a\Sigma)^*
+(Y_i\Sigma)(Y_i\partial_a^2\Sigma)^*\right]\ .
\label{dvs}
\eea
For the fermion induced scalar operators ${\cal O}_F(\Sigma)$ in the 
Lagrangian we have to compute
\be
\label{dfs}
\delta_F  {\mathrm Tr}[{\cal M}_S^2(\Sigma)]=
4c_F f^4 h_1^2\sum_a \partial_a^2
{\mathrm Tr} 
[\Sigma_1\epsilon\Sigma_1^T\Sigma_1^*\epsilon\Sigma_1^\dagger]\ ,
\ee
which we refrain from expanding further in terms of $\partial_a\Sigma_1$ 
and $\partial_a^2\Sigma_1$ because the final expression is too lengthy.

Therefore, we need to compute derivatives of $\Sigma$ with respect to the 
real scalar fields $\pi_a$ in $\Pi=\sum_a \pi_a X_a$, evaluated at 
$\pi_a=0$. Once we have 
eq.~(\ref{good}), such derivatives of $\Sigma$ are simple to compute. One 
gets
\bea
\partial_a\left.\Sigma\right|_0 & = & {2i\over f}e^{i\langle\Pi\rangle/f}  
X_a e^{i\langle\Pi\rangle/f} \Sigma_0\ ,\nonumber\\
\partial^2_a\left.\Sigma\right|_0 & = &- {4\over f^2}e^{i\langle\Pi\rangle/f}
  X_a X_a e^{i\langle\Pi\rangle/f} \Sigma_0\ ,
\label{der}
\eea
and their conjugates
\bea
\partial_a\left.\Sigma^*\right|_0 & = & 
-{2i\over f}e^{-i\langle\Pi^T\rangle/f}  
X_a^* e^{-i\langle\Pi^T\rangle/f} \Sigma_0\ ,\nonumber\\
\partial^2_a\left.\Sigma^*\right|_0 & = &
- {4\over f^2}e^{-i\langle\Pi^T\rangle/f}
  X_a^* X_a^* e^{-i\langle\Pi^T\rangle/f} \Sigma_0\ .
\label{derc}
\eea 

From this point onwards we drop the brackets in $\langle\Pi\rangle$  
as we are going to interpret ${\mathrm Tr}[{\cal M}^2(\Sigma)]$ as a source 
of new operators to be added to the Lagrangian from 
the beginning. To compute this trace, the second derivatives in (\ref{der}) 
and (\ref{derc}) will always be summed in $a$. Then notice that, using the 
fact that the $X_a$ are the broken generators, one has
\be
\sum_a X_a X_a = C_{SU(5)}(5)-C_{SO(5)}(5)= {14\over 
5} I_5\ , 
\label{xaxa}
\ee
(where the $C$'s are the quadratic Casimir operators for the fundamental 
representations) and so, the terms in 
(\ref{dvs}) and (\ref{dfs}) containing 
$\partial_a^2\Sigma$ and $\partial_a^2\Sigma_1$ are trivial to compute. In 
fact, all such terms generate operators proportional to ${\cal 
O}_{V_i}(\Sigma)$, ${\cal O}_{Y_i}(\Sigma)$ or ${\cal O}_F(\Sigma)$. The 
rest 
of the terms are more interesting and a bit harder to compute. The 
simplest way to proceed is to notice the following two 
identities (derived in Appendix~C), valid 
for generic $5\times 5$ matrices $Y$ and $Z$:
\bea
\sum_a {\mathrm Tr}[X_a YX_a^* Z]&=&
{1\over 2} {\mathrm Tr}[YZ^T]-{1\over 5}{\mathrm Tr}[YZ]
+{1\over 2}{\mathrm Tr}[Y\Sigma_0^*]{\mathrm Tr}[Z\Sigma_0]\ 
,\nonumber\\
\sum_a {\mathrm Tr}[X_a YX_a Z]&=&
{1\over 2} {\mathrm Tr}[Y]{\mathrm Tr}[Z]
-{1\over 5}{\mathrm Tr}[YZ]+{1\over 2}
{\mathrm Tr}[Y\Sigma_0Z^T \Sigma_0^*]\ .
\label{ident}
\eea
[Note in particular that eq.~(\ref{xaxa}) agrees with the second formula 
above, for the particular case $Y=Z=I_5$.] By using these 
identities one 
can easily show that $\delta_V {\mathrm Tr}[{\cal M}_S^2(\Sigma)]$ in 
(\ref{dvs}) gives $\Sigma$-operators proportional to ${\cal 
O}_{V_i}(\Sigma)$ and ${\cal O}_{Y_i}(\Sigma)$ or field-independent. Most 
of 
the contributions from $\delta_F {\mathrm Tr}[{\cal M}_S^2(\Sigma)]$ in 
(\ref{dfs}) are also proportional to ${\cal O}_F(\Sigma)$ but a few give a 
different operator. These are
\be
4c_Ff^4h_1^2\sum_a {\rm Tr}\left[\partial_a\Sigma_1 \epsilon
\Sigma_1^T(\partial_a\Sigma_1^*\epsilon\Sigma_1^\dagger
+\Sigma_1^*\epsilon\partial_a\Sigma_1^\dagger)
\right]+{\rm h.c.}\ ,
\ee
which produce terms proportional to
\be
{\cal O}_S(\Sigma)\equiv {\mathrm Tr}[\Sigma_1^\dagger\Sigma_1]=
\Sigma_{ix}{\Sigma^{ix}}^*\ .
\ee
This looks like a new type of operator one has to add to 
the Littlest Higgs 
Lagrangian from the start.

Before analyzing ${\cal O}_S(\Sigma)$ by expanding it in powers of the 
physical fields $H$ and $\Phi$ (which we leave for the next section) we 
have to discuss the following two issues. The first is to calculate the 
contribution of ${\cal O}_S(\Sigma)$ itself to ${\mathrm Tr}[{\cal 
M}^2(\Sigma)]$. It is in fact quite easy to see that such contribution,
\be
\label{dss}
\delta_S  {\mathrm Tr}[{\cal M}_S^2(\Sigma)]\propto \sum_a
{\mathrm Tr}\left[(\partial_a^2\Sigma_1)\Sigma_1^\dagger
+2(\partial_a\Sigma_1)(\partial_a\Sigma_1^\dagger)
+\Sigma_1(\partial_a^2\Sigma_1^\dagger)\right]\ ,
\ee
is proportional to ${\cal O}_S(\Sigma)$ so that no other new operators are 
generated and we can stop at this point. The second issue is to check that 
${\cal O}_S(\Sigma)$ does indeed respect the symmetries of the low-energy 
effective theory, in particular the $SU(3)_i$ symmetries that are 
all-important to guarantee the lightness of the Higgs. We do this in the 
following section.

\section{Scalar Operators Revisited}

When $g_1=g_1'=0$ the Lagrangian is invariant under $SU(3)_1\times 
SU(2)_2$ 
with $SU(3)_1$ global and $SU(2)_2$ local, with $\Sigma\rightarrow 
U\Sigma U^T$ and
\be
\label{type1}
U=\left(\begin{array}{cc}
U_1 &0\\
0&V_2
\end{array}\right)\ ,
\ee
where $U_1$ is a $3\times 3$ $SU(3)$ matrix and $V_2$ a $2\times 2$ 
$SU(2)$ 
matrix. We will call this a type-1 symmetry transformation. Under the 
global $SU(3)_1$, the Higgs shifts by a constant and 
this forbids a mass term for it. When $g_2=g'_2=0$ the Lagrangian is 
invariant 
under $SU(2)_1\times SU(3)_2$ with $SU(3)_2$ global and $SU(2)_1$ local, 
as before, but with 
\be
\label{type2}
U=\left(\begin{array}{cc}
V_1 &0\\
0&U_2
\end{array}\right)\ ,
\ee
where $U_2$ is a $3\times 3$ $SU(3)$ matrix and $V_1$ a $2\times 2$ 
$SU(2)$
matrix (type-2 symmetry transformation). Again, invariance under the 
global 
$SU(3)_2$ forbids a mass term 
for the Higgs. Generating a non-zero Higgs mass requires both type-1 and 
type-2 couplings simultaneously and quadratically divergent 
one-loop radiative corrections do not 
generate such mass term (only the softer logarithmically divergent 
corrections can generate it).

Under type-1 symmetry transformations (\ref{type1}), $\Sigma_1$ transforms 
simply as
\be
\Sigma_1\rightarrow U_1\Sigma_1 V_2^T\ ,
\ee
and ${\mathrm Tr}[\Sigma_1^\dagger\Sigma_1]$ is obviously invariant. Under 
the type-2 transformations, (\ref{type2}), $\Sigma_1$ does not 
transform in a simple way under $SU(3)_2$ and ${\mathrm 
Tr}[\Sigma_1^\dagger\Sigma_1]$ is not invariant under that global 
symmetry. Of course invariance under the local $SU(2)_1$ is still 
maintained [$SU(2)_1$ is in fact a subgroup of $SU(3)_1$].

After this simple discussion several facts become obvious. First, using 
$\Sigma_1$ it is straightforward to construct other operators that are 
type-1 invariant, for instance
\be
{\cal O}_n\equiv{\mathrm Tr}[(\Sigma_1^\dagger\Sigma_1)^n]\ .
\ee
In addition to these, one can make use of the $2\times 
2$ antisymmetric tensor $\epsilon$ to get new $SU(2)$ invariant 
combinations and write type-1 operators like
\be
\label{One}
{\cal O}_n^\epsilon\equiv{\mathrm Tr} 
[(\Sigma_1\epsilon\Sigma_1^T\Sigma_1^*\epsilon\Sigma_1^\dagger)^n]\ .
\ee
For $n=1$ this corresponds to the scalar operator induced by fermion 
loops.
Other combinations give mixed operators like
\be
{\cal O}_m\equiv{\mathrm Tr}
[(\Sigma_1\epsilon\Sigma_1^T)
(\Sigma_1^*\Sigma_1^T)(\Sigma_1^*\epsilon\Sigma_1^\dagger)]\ ,
\ee
and so on.

Finally, it is clear that type-1 and type-2 symmetry transformations are 
quite similar and therefore one can define another $3\times 2$ submatrix 
\be
\Sigma_2=(\Sigma_{\hat i\hat x})\ ,
\ee
with $\hat i=\{3,4,5\}$ and $\hat x=\{1,2\}$. Under type-2 
symmetry transformations we have
\be
\Sigma_2\rightarrow U_2\Sigma_2V_1^T\ ,
\ee 
in such a way that operators invariant under type-2 transformations are 
trivial to construct, simply replacing $\Sigma_1\rightarrow\Sigma_2$ in 
all the operators above.

The scalar operators induced by loops of gauge bosons,
\bea
{\cal O}_{V_i}(\Sigma)&\equiv &\sum_\alpha{\mathrm 
Tr}[(Q_i^{\alpha}\Sigma)(Q_i^{\alpha}\Sigma)^*]\ ,\nonumber\\
{\cal O}_{Y_i}(\Sigma)&\equiv & 
{\mathrm Tr}[(Y_i\Sigma)(Y_i\Sigma)^*]\ ,
\eea
can in fact be rewritten in a simpler way in terms of the former operators 
for $\Sigma_1$ and $\Sigma_2$. One gets
\bea
{\cal O}_{V_1}(\Sigma)&= &{1\over 2}-{1\over 4}{\mathrm 
Tr}[\Sigma_2^\dagger\Sigma_2]\ ,\nonumber\\
{\cal O}_{Y_1}(\Sigma)&=& {3\over 10}-{1\over 4}{\mathrm
Tr}[\Sigma_2^\dagger\Sigma_2]\ ,
\eea
and
\bea
{\cal O}_{V_2}(\Sigma)&= &{1\over 2}-{1\over 4}{\mathrm
Tr}[\Sigma_1^\dagger\Sigma_1]\ ,\nonumber\\
{\cal O}_{Y_2}(\Sigma)&= &{3\over 10}-{1\over 4}{\mathrm
Tr}[\Sigma_1^\dagger\Sigma_1]\ ,
\eea
The scalar operator induced by fermion loops, ${\cal O}_F(\Sigma)={\cal 
O}_1^\epsilon$, see (\ref{One}), can be written simply as
\be
{\cal O}_F(\Sigma)=-2\ {\mathrm Det}[\Sigma_1^\dagger \Sigma_1]\ .
\ee 
In fact, all operators that respect the type-1,2 symmetries 
can be expressed in terms of 4 invariant 
quantities, the trace and determinant of the $2\times 2$ Hermitian 
operators ${\cal H}_1 Â\equiv \Sigma_1^\dagger\Sigma_1$ and ${\cal H}_2 
\equiv \Sigma_2^\dagger\Sigma_2$, which can be considered the building 
blocks of such scalar operators.  
For instance
\bea
{\mathrm Tr}[{\cal H}_{1,2}^2]&=&{\mathrm Tr}[{\cal H}_{1,2}]^2-2{\mathrm 
Det}[{\cal H}_{1,2}]\ 
,\nonumber\\
{\mathrm Tr}[{\cal H}_{1,2}\epsilon 
{\cal H}_{1,2}^T{\cal H}_{1,2}^T\epsilon]&=&-{\mathrm 
Tr}[{\cal H}_{1,2}]\ {\mathrm Det}[{\cal H}_{1,2}]\ 
,
\eea
and so on.

Let us  now expand these 4 invariants in terms of the physical fields: the 
Higgs doublet $H$ and the triplet $\Phi$. Up to fourth order in the scalar 
fields, we obtain the following expansions
\bea
\label{Trexp}
{\mathrm Tr}[\Sigma_{1,2}^\dagger\Sigma_{1,2}]&=&2-{4\over f^2}{\mathrm 
Tr}[\Phi^\dagger\Phi]
\pm{2i\over f^3}\left[H^*\Phi H^\dagger-{\mathrm h.c.}\right]
-{1\over f^4}|H|^4\nonumber\\
&&+{16\over 3f^4}{\mathrm Tr}[\Phi^\dagger\Phi\Phi^\dagger\Phi]
+{16\over 3f^4}H^*\Phi\Phi^\dagger H^T+...
\eea
\bea
\label{Detexp}
{\mathrm Det}[\Sigma_{1,2}^\dagger\Sigma_{1,2}]&=&1-{4\over f^2}{\mathrm 
Tr}[\Phi^\dagger\Phi]
\pm{2i\over f^3}\left[H^*\Phi H^\dagger-{\mathrm h.c.}\right]
-{1\over f^4}|H|^4\nonumber\\
&&-{8\over 3f^4}{\mathrm Tr}[\Phi^\dagger\Phi\Phi^\dagger\Phi]
+{8\over f^4}({\mathrm Tr}[\Phi^\dagger\Phi])^2
+{16\over 3f^4}H^*\Phi\Phi^\dagger H^T+...
\eea
(with the $+$ sign for the type-1 case and the $-$ sign for the type-2 case).
Much of this structure is in fact dictated by the global symmetries from 
the first terms in the expansion [although not completely: notice that 
the quartic couplings for $\Phi$ do differ between (\ref{Trexp}) and 
(\ref{Detexp})] 
so that in retrospective it is not surprising that scalar loops generate 
operators that were already induced by fermion or gauge boson loops.

We see that, provided the coefficients in front of these operators have
the appropriate sign, the triplet $\Phi$ gains a heavy mass of order $f$. 
Below that threshold, one can integrate out this triplet. If the 
only operators present are type-1 or type-2, one simply gets
\be
\Phi_{ij}= \pm {i\over 2f} H_i H_j + f {\cal O}(H^4/f^4)\ ,
\ee
and substituting back in (\ref{Trexp}) or (\ref{Detexp}) we see that there 
is a cancellation and the Higgs quartic coupling is zero. In the presence 
of both type-1 and type-2 operators however, this is not the case. 
From
\bea
\lambda_1{\cal O}_1(\Sigma)+\lambda_2{\cal O}_2(\Sigma)&=&
(\lambda_1+\lambda_2)f^2{\mathrm 
Tr}[\Phi^\dagger\Phi]+{i\over 2}(\lambda_1-\lambda_2)
f[H^*\Phi H^\dagger-h.c.]\nonumber\\
&+&{1\over 4}(\lambda_1+\lambda_2)|H|^4+...\ ,
\label{lambda}§
\eea
one obtains now
\be
\Phi_{ij}=\pm {(\lambda_1-\lambda_2)\over (\lambda_1+\lambda_2)}H_iH_j + 
{\cal O}(H^4)\ ,
\ee
and substituting back in (\ref{lambda}) one gets a quartic coupling 
$(\lambda/4)|H|^4$ satisfying
\be
{1\over\lambda}={1\over\lambda_1}+{1\over \lambda_2}\ .
\ee
Once again we see how both type-1 and type-2 couplings must be present 
simultaneously, this time to generate a non-zero Higgs quartic coupling.

Due to the symmetry structure of the Littlest Higgs model the 
$\Sigma$-operator induced by scalar loops turned out to be similar to the 
operators induced by gauge boson loops. In other LH models this does not 
need to be the case. In Appendix~B we give an example, based on the 
$SU(6)/Sp(6)$ coset structure, in which scalar loops generate genuinely 
new operators. Another example is the ``Simplest'' Little Higgs model of 
ref.~\cite{Simplest} in which fermion and gauge boson loops do not induce 
any scalar operators. These have to be introduced by hand and once this is 
done they do get renormalized by scalar loops.

\section{Implications}

Consider a simple tree-level scalar potential for the Littlest Higgs of 
the form
\be
V_0/f^4=
\kappa_1 {\mathrm Tr}[\Sigma_1^\dagger\Sigma_1]+
\kappa_2 {\mathrm Tr}[\Sigma_2^\dagger\Sigma_2]+
\kappa'_1 {\mathrm Det}[\Sigma_1^\dagger\Sigma_1]+
\kappa'_2 {\mathrm Det}[\Sigma_2^\dagger\Sigma_2]\ .
\label{Vtree}
\ee
In principle, operators of higher order in $\Sigma_{1,2}$ could be added
but for our purposes it will be enough to keep these four.
The constants $\kappa_{1,2}$ and $\kappa'_{1,2}$ are unknown (they depend 
on 
the UV completion of the model). One can nevertheless estimate their 
typical size simply by looking at the radiative contributions they receive 
from loops of light particles. The most important correction comes from
the quadratically divergent piece $\Lambda^2\  {\mathrm Str} M^2/(32\pi^2)$. 
In this way, and using the technique explained in sect.~2 we obtain
\bea
\delta\kappa_1&=&-\left[{3\over 
16}(g_2^2+{g'}_2^2)+12\kappa_1-4\kappa'_1\right]
{\Lambda^2\over (4\pi f)^2}\ ,\nonumber\\
\delta\kappa_2&=&-\left[{3\over 
16}(g_1^2+{g'}_1^2)+12\kappa_2-4\kappa'_2\right]
{\Lambda^2\over (4\pi f)^2}\ ,\nonumber\\
\delta\kappa'_1&=&-\left[6h_1^2+20\kappa'_1\right]{\Lambda^2\over (4\pi 
f)^2}\ ,\nonumber\\
\delta\kappa'_2&=&-20\kappa'_2{\Lambda^2\over (4\pi f)^2}\ ,
\label{main}
\eea
and here one can substitute $\Lambda=4\pi f$ for the NDA estimate. 
The interest of this calculation is the following. 
Note that it would not be natural to expect that the unknown constants 
$\kappa_{1,2}$ and $\kappa'_{1,2}$ take numerical values much smaller than 
the 
one-loop contributions displayed in (\ref{main}). Concerning the gauge 
boson and fermion loop contributions, they can already  cause problems 
with the naturalness of electroweak breaking in these models \cite{CEH}: 
phenomenology often requires values for $\kappa_{1,2}$ and $\kappa'_{1,2}$  
which are indeed significantly smaller than such one-loop corrections. 

Eq.~(\ref{main}) includes also the new contributions from scalar loops, 
which have not been obtained in the literature before and have important 
implications. As is clear from (\ref{main}) this scalar contribution is 
even more problematic than the rest: no matter what tree-level value for 
$\kappa_{1,2}$ and $\kappa'_{1,2}$ one starts with (provided it is not 
zero), loop contributions tend to generate a value which is one order of 
magnitude larger. The problem with this is not simply that one-loop 
corrections are larger than the tree level result (this does not mean 
necessarily that the perturbative expansion is breaking down). For 
instance, one-loop corrections might involve some coupling that is 
significantly larger than the couplings entering the tree-level result 
so that they overcome the loop suppression (the corrections to the Higgs 
mass in the 
MSSM \cite{MSSMmh} are a famous example). The problem with (\ref{main}) is 
that one-loop corrections involve precisely the same coupling entering at 
tree-level. In other words, these scalar couplings get 
multiplicatively renormalized as
\be
\kappa\rightarrow\kappa\left[1-{\cal N}{\Lambda^2\over(4\pi f)^2}\right]\ 
,
\ee
with large ${\cal N}$ (${\cal N}=12$ or 20 in this particular model), so 
that perturbative calculability is really lost.

The root of this pathological behaviour is the NDA 
substitution $\Lambda=4\pi f$, and we are thus lead to conclude that the 
self-consistency of the whole non-linear sigma model approach requires a 
lower cutoff, satisfying at least
\be
\Lambda \simlt {4\pi f \over\sqrt{20}}\ ,
\label{mainL}
\ee
which, furthermore, probably cannot be saturated.
This is in agreement with general results in technicolor theories: for 
instance, ref.~\cite{sqrtN} shows that the chiral perturbation theory that 
describes the low-energy behaviour of technicolor theories breaks down at 
$4\pi f /\sqrt{{\cal N}_{tf}}$, where ${\cal N}_{tf}$ is the number of 
technifermions. Notice 
however that the result (\ref{mainL}) arises from purely low-energy 
considerations about the consistency of the effective theory approach 
without reference to the possible UV completion of the LH model.

If we interpret this UV cutoff as the scale at which the theory becomes 
strongly interacting (usually taken to be $4\pi f$ instead) then, the 
implication for model building of the stricter upper bound (\ref{mainL}) 
is that it is not possible to keep $\Lambda=10$ TeV and $f=1$ TeV. In 
fact, if for naturalness one keeps $f=1$ TeV, the hierarchy one would be 
able to explain with this kind of Little Higgs theories will be much milder: 
$\Lambda\simeq 3$ TeV. Conversely, if one insists in keeping 
$\Lambda \simeq 10$ TeV, the scale $f$ is not stable at $1$ TeV but rather 
pushed up to $f\simeq 3.5$ TeV. While this can 
in fact be welcome to avoid problems with electroweak precision tests 
\cite{Csaki,EWPT}, it would generically be a disaster for the naturalness 
of 
electroweak breaking \cite{CEH}.

A different possibility is that the scale $4\pi f/\sqrt{20}$ signals the 
appearance of new degrees of freedom/resonances (besides the 
pseudo-Goldstones included already in the low-energy effective theory)  
lighter than the scale of strong dynamics at $4\pi f$. 
Such possibility can in fact be quite generic and has been discussed 
previously in the context of Little Higgs models \cite{reson}. In this 
respect, our argument in favor of such new states is purely bottom-up, 
with no reference to a particular UV completion but rather coming from 
the consistency of the low-energy effective theory. In this sense our 
results are 
complementary to unitarity arguments \cite{unitarity} which also point 
towards a cutoff scale lower than the NDA estimate. Typically, our bounds on 
$\Lambda$ are somewhat stronger than those coming from unitarity  [{\it 
e.g.}~for the $SU(5)/SO(5)$ model  we get $\Lambda\simlt 
4\pi f/\sqrt{20}\simeq 2.81 f$ vs.~the unitarity bound  $\Lambda_U\simlt 
3.17 f$  or $\Lambda\simlt 4\pi f/\sqrt{24}\simeq 2.57 f$ 
vs.~$\Lambda_U\simlt 3.68 f$ for the $SU(6)/Sp(6)$ model \cite{unitarity}].

The size of the factor ${\cal N}$ seems to grow with the size of the 
global groups involved. In Appendix~C we have performed the exercise of 
calculating this bound for a $SU(N)/SO(N)$ non-linear sigma model to see 
how ${\cal N}$ scales with $N$ finding ${\cal N}=4N$. This seems then to 
favour LH models with small group structures. We have analyzed as a 
further example the ``Simplest'' LH model \cite{Simplest}, based on 
$[SU(3)\times U(1)]^2/[SU(2)\times U(1)]^2$, using the same techniques 
finding 
\be
\Lambda\simlt {4\pi f_1 f_2\over \sqrt{3} f}\leq {4\pi f\over 2\sqrt{3}}
\ ,
\ee
where $f_1$ and $f_2$ are the two order parameters of the 
spontaneous symmetry breaking in this model, satisfying $f_1^2+f_2^2=f^2$. 
The second inequality follows from this constraint and we therefore arrive 
at ${\cal N}\geq 12$, which is indeed smaller than for larger groups but 
still significant.

Before moving on to other implications we should emphasize that the 
bounds on the cutoff we have derived by looking at the quadratically 
divergent corrections induced by scalar couplings are independent of how 
we choose to parametrize the scalar operators. For instance, instead of 
using the operators $\{{\rm Tr}[\Sigma_1^\dagger\Sigma_1],
{\rm Tr}[\Sigma_2^\dagger\Sigma_2], {\rm Det}[\Sigma_1^\dagger\Sigma_1],
{\rm Det}[\Sigma_2^\dagger\Sigma_2]\}$ we could have chosen any other 
linear combination of them (provided the new four operators are 
independent). In general, starting with a tree-level potential
\be
V_0/f^4=\sum_{\alpha=1}^n \kappa_\alpha{\cal O}_\alpha(\Sigma)\ ,
\ee
we could change to a new basis of operators $\{{\cal O}'_\alpha(\Sigma)\}$
with
\be
{\cal O}'_\alpha(\Sigma)=A_{\alpha\beta}{\cal O}_\beta(\Sigma)\ ,
\ee
where the matrix $A$ is non-singular (we are assuming repeated indices 
are summed over). In this new basis the tree-level potential is
\be
V_0/f^4=\sum_{\alpha=1}^n \kappa'_\alpha{\cal O}'_\alpha(\Sigma)\ ,
\ee
with
\be
\kappa'_\alpha=\kappa_\beta A^{-1}_{\beta\alpha}\ .
\ee
In the original basis, the quadratically divergent correction induced by 
these scalar operators takes the form
\be
\label{R}
\delta V/f^4=-{\Lambda^2\over (4\pi f)^2}\kappa_\alpha 
R_{\alpha\beta}{\cal 
O}_\beta(\Sigma)\ ,
\ee
where $R$ is a numerical matrix (not normal in general). In the new basis 
the form of $\delta V/f^4$ is the same with primed quantities and the new 
matrix $R'$ is related to the original $R$ by a similarity transformation:
\be
\label{simil}
R'_{\alpha\beta}=A^{-1}_{\alpha\gamma}R_{\gamma\rho}A_{\rho\beta}\ .
\ee
Notice that we have implicitly assumed
that the bases are complete, in the sense that no new operators appear in 
$\delta V$ that were not present already in $V_0$. We also assume that $R$ 
is of rank $n$. If this were not the case and ${\rm Rank}(R)=n'<n$ then we 
could have started with a smaller basis with just $n'$ operators.
Therefore, the matrix $R$ can be diagonalized and concerning the 
bound on the cutoff scale $\Lambda$ we are interested in its largest 
eigenvalue (in absolute value). It is clear then from the above discussion 
that the bound does not depend on what basis is chosen for the analysis 
because a similarity transformation like (\ref{simil}) leaves the 
eigenvalues of $R$ invariant. 

In the Littlest Higgs model, the matrix $R$ can be read off directly from 
eq.~(\ref{main}):
\be
R=4\left(\begin{array}{cccc}
3 & º0 & -1 & 0\\
0 & 3 & 0 & -1\\
0 & 0 & 5 & 0\\
0 & 0 & 0 & 5
\end{array}\right)\ ,
\ee
and has the eigenvalues $\{12,12,20,20\}$ so that the 
bound on $\Lambda$ is indeed $\Lambda\simlt 4\pi f/\sqrt{20}$.

Finally, the new scalar contributions we have discussed might have 
implications for the important issue of vacuum alignment \cite{va0,va}.
Expanding in physical fields the scalar potential (\ref{Vtree}), with couplings 
corrected by one-loop quadratic divergences as given in (\ref{main}), we 
get
\bea
V&=&4f^2{\rm Tr}[\Phi^\dagger\Phi]
\left\{{\over}-(\kappa_1+\kappa_2+\kappa_1'+\kappa_2')
\right.\\
&+&\left.{\Lambda^2\over (4\pi f)^2}\left[{3\over 
16}(g_1^2+g_2^2+{g'_1}^2+{g'_2}^2)+6h_1^2
+12(\kappa_1+\kappa_2)+16(\kappa'_1+\kappa_2')
\right]\right\}+...\nonumber
\eea
It is easy to choose the $\kappa_i$ and $\kappa_i'$ couplings in such a 
way that the mass of the triplet $\Phi$ is positive, for instance by  
choosing them all negative. Take into account that $\Lambda$ is bounded 
precisely in such a way that the scalar one-loop divergent piece cannot 
overcome the tree level contribution. 
It might seem that this last fact precludes the 
scalar radiative correction from having any effect in vacuum alignment. 
This is not so. For instance, assume that gauge and 
fermion couplings are negligible compared with the scalar  $\kappa_i$ 
and $\kappa_i'$ couplings so that 
\be
V\simeq -4f^2{\rm Tr}[\Phi^\dagger\Phi]\left\{{\over}(\kappa_1+\kappa_2)
\left[1-12{\Lambda^2\over (4\pi f)^2}\right]
+(\kappa'_1+\kappa'_2)\left[1-16{\Lambda^2\over (4\pi 
f)^2}\right]\right\}+...
\label{vaex}
\ee
Next take $\kappa_1+\kappa_2>0$ and $\kappa_1'+\kappa_2'<0$ but with 
$\kappa_1+\kappa_2+\kappa_1'+\kappa_2'<0$ (so 
that the chosen vacuum is stable at tree level). Scalar loop corrections 
in (\ref{vaex}) tend to reduce in absolute value the $\kappa_1+\kappa_2$
and $\kappa_1'+\kappa_2'$ contributions but the effect is stronger for 
this latter (positive) piece. It is then possible that for some $\Lambda$ 
(well below
the upper bound $4\pi f/\sqrt{20}$) the negative contribution from 
$\kappa_1+\kappa_2$ dominates and destabilizes the vacuum.
In any case, such possibilities and choices of couplings should be 
discussed in the context of a UV complete model.

\section{Conclusions}

Little Higgs models protect the mass of the Higgs boson from one-loop 
quadratically divergent corrections by making it a pseudo-Goldstone of 
several global symmetries broken in a collective way at some scale 
$f\sim 1$~TeV. The models try to 
stabilize in this manner the little hierarchy between the electroweak 
scale and the 10 TeV scale, where new physics (presumably strongly 
coupled) appears. Although it is not clear whether a fully satisfactory
model exists (the models in the literature are either more finetuned 
than naively thought \cite{CEH} or have problems with precision 
electroweak tests \cite{Csaki,EWPT} or both) the idea is interesting in 
principle.

These models predict the existence of a set of new particles with masses 
of order $f$ which fill out multiplets of the global symmetries 
together with the SM particles and are responsible for cancelling (at one 
loop) the dangerous quadratic divergences that affect the SM Higgs boson 
mass. Only this mass (or perhaps that of a second Higgs doublets in some 
LH versions) is protected from one-loop quadratic divergences. The scalar 
potential receives the usual quadratically divergent contribution
\be
\label{Vquad}
\delta V={\Lambda^2\over 32\pi^2}{\rm Str}M^2\ ,
\ee
which is in fact crucial for the phenomenological viability of the models:
it generates scalar operators that give masses of order $f$ to the  
particles beyond the SM ones introduced in these models, a Higgs quartic 
coupling, etc. In previous literature the contributions of gauge boson and 
fermion loops to the supertrace in (\ref{Vquad}) have been computed and 
discussed and the size of the resulting scalar operators is estimated 
using the Naive Dimensional Analysis rule $\Lambda\simeq 4\pi f$ (which is 
roughly 10~TeV for $f\simeq 1$~TeV). These radiatively induced scalar 
operators are then included in the Lagrangian from the very beginning: 
they could clearly be present after integrating out the new physics at 
$\Lambda$. Once this is done there will be also a scalar contribution to 
the supertrace in  (\ref{Vquad}) and the main purpose of this paper has 
been to compute this new contribution and discuss what impact it might 
have for the phenomenology of LH models.
In calculating this new scalar contribution to the effective potential we 
have used with advantage a technique proposed recently in \cite{ELR} which 
greatly simplifies the task.

We have found that, depending on the structure of global symmetries 
respected by the gauge sector and the fermion sector, it is possible that 
scalar loops generate truly new operators not induced before by loops of 
gauge bosons or fermions. Such is the case for some versions of the 
$SU(6)/Sp(6)$ LH model although not for the Littlest Higgs model. 
More importantly, we have found that in general scalar loops renormalize 
the existent scalar operators multiplicatively in the form 
\be
{\cal O}_S\rightarrow{\cal O}_S\left[1-{\cal N}{\Lambda^2\over 
(4\pi f)^2}\right]\ ,
\ee
where ${\cal N}$ is sizeable [{\it e.g.}~${\cal N}=20$ in the Littlest 
Higgs model, ${\cal N}=24$ in the version of the $SU(6)/Sp(6)$ model 
discussed in Appendix~B, ${\cal N}\geq 12$ in the ``Simplest'' LH model, 
etc.]. 
Consistency of the non-linear effective theory approach demands 
\be
\Lambda\simlt {4\pi f\over\sqrt{{\cal N}}}\ ,
\ee
which is in general significantly lower than the NDA estimate $4\pi 
f\simeq 10$~TeV. This can hinder the naturalness of the separation between 
$f\simeq 1$ TeV and $\Lambda\simeq 10$ TeV, causing further difficulties 
in LH model building. Alternatively, it may signal the appearance of new 
resonances well below 10~TeV. These may be a concern for electroweak 
precision tests or may offer new experimental handles on LH models at the LHC.

\section*{A. Log Corrections in the ${\bma{SU(5)/SO(5)}}$ Model}
\setcounter{equation}{0}
\renewcommand{\theequation}{A.\arabic{equation}}

The same technique used in the main text can be applied to the calculation 
of the logarithmically divergent corrections to the effective potential,
\be
\delta_L V=-{\log\Lambda^2\over 64\pi^2}{\rm Str}M^4\ .
\ee
Just for illustrative purposes, in this appendix we calculate the 
scalar contributions to $\delta_L V$ coming from the following part of the  
tree-level scalar potential of eq.~(\ref{Vtree})
\be
\delta V_0/f^4=
\kappa_1 {\mathrm Tr}[\Sigma_1^\dagger\Sigma_1]+
\kappa_2 {\mathrm Tr}[\Sigma_2^\dagger\Sigma_2]\ ,
\label{Vtree0}
\ee
{\it i.e.}~we will write down the corrections proportional to 
$\kappa_1^2$, $\kappa_2^2$ and $\kappa_1\kappa_2$. Because 
the latter corrections break both global $SU(3)_i$ symmetries 
(putting in communication sectors with different types of
global symmetries) the operators proportional to 
$\kappa_1\kappa_2$ generated in this way cannot be 
expressed in terms of $\Sigma_1$ and $\Sigma_2$ only. It is then convenient to 
split these matrices in the following way:
\be
\Sigma_1=\left(\begin{array}{c}
\Theta \\
\sigma_1
\end{array}\right)\ ,\;\;\;\;
\Sigma_2=\left(\begin{array}{c}
\sigma_2^T\\
\Theta^T 
\end{array}\right)\ ,
\ee
where $\Theta$ is a $2\times2$ matrix and $\sigma_i$ are 2-dimensional 
vectors. Under $SU(2)_1\times SU(2)_2$ gauge transformations, with 
matrices $V_1$ and $V_2$, these quantities change as
\bea
\Theta & \rightarrow & V_1\Theta V_2^T\ ,\nonumber\\
\sigma_1 & \rightarrow & \sigma_1 V_2^T\ ,\nonumber\\
\sigma_2 & \rightarrow & V_1\sigma_2 \ .
\label{transf}
\eea

Using the previous decomposition one gets
\bea
\delta_{L(1,2)} V&=& \left.-{\kappa_1\kappa_2 \over 4\pi^2}
f^4\log\Lambda^2\right\{8+2\
{\rm Tr}[\Sigma_1^\dagger\Sigma_1]{\rm Tr}[\Sigma_2^\dagger\Sigma_2]
-4\ ({\rm Tr}[\Sigma_1^\dagger\Sigma_1]+{\rm 
Tr}[\Sigma_2^\dagger\Sigma_2])
\nonumber\\
&+&2\ {\rm Tr}[\Theta^\dagger\Theta]^2
+9\ {\rm Tr}[\Theta^\dagger\Theta\Theta^\dagger\Theta]
-7\ {\rm Tr}[\Theta^\dagger\Theta]
+7\ (\sigma_1\Theta^\dagger\Theta\sigma_1^\dagger+
\sigma_2^\dagger\Theta\Theta^\dagger\sigma_2)\nonumber\\
&+&\left.
{7\over 2}
(\Sigma_{33}^*\sigma_1\Theta^\dagger\sigma_2+{\rm h.c.})\right\}\ .
\eea
One can check, from (\ref{transf}) and the fact that $\Sigma_{33}$ is an 
$SU(2)_1\times SU(2)_2$ singlet, that all these operators are 
gauge-invariant. As expected, an expansion of this correction in powers of 
the physical fields shows a log-divergent contribution to the mass of the 
Higgs boson.

For comparison, we also write down the corrections proportional to 
$\kappa_a^2$, which do not provide such a contribution to the Higgs mass:
\be
\delta_{L(a)} V= -{\kappa_a^2\over 8\pi^2}
f^4\log\Lambda^2\left\{4\ {\rm Tr}[\Sigma_a^\dagger\Sigma_a]^2
+9\ {\rm Tr}[\Sigma_a^\dagger\Sigma_a\Sigma_a^\dagger\Sigma_a]
-17\ {\rm Tr}[\Sigma_a^\dagger\Sigma_a]\right\},
\ee
with $a=1,2$. In a similar way, one could compute the logarithmically 
divergent corrections coming from fermion, gauge boson and other scalar loops.

\section*{B. Scalar Loops in the ${\bma{SU(6)/Sp(6)}}$ Model}
\setcounter{equation}{0} 
\renewcommand{\theequation}{B.\arabic{equation}}

This model \cite{SU6Sp6} is based on an $SU(6)/Sp(6)$ nonlinear sigma 
model, with the spontaneous breaking of $SU(6)$ down to $Sp(6)$ 
produced by the vacuum expectation value of a $6\times 6$ antisymmetric 
matrix field $\Phi$.  We follow \cite{GWS} and choose 
\beq 
\langle\Phi\rangle\equiv\Sigma_0 =
\left(\begin{array}{cc}  
\begin{array}{cc} 
  & I_2 \\ -I_2 &  
\end{array} &  \\
 & \begin{array}{cc}
  & 1 \\ -1 & 
\end{array}
\end{array}\right)\ . 
\label{vevsp} 
\eeq 
This breaking of the global $SU(6)$ symmetry produces 14 Goldstone bosons 
which include the Higgs doublet field. As usual, these Goldstone bosons 
can be parametrized through the nonlinear sigma model field 
\beq 
\Sigma = e^{i \Pi/f} \Sigma_0 e^{i \Pi^T/f}\ , 
\label{Sigmasp} 
\eeq 
with $\Pi = \sum_a \pi^a X^a$, where $\pi_a$ are the Goldstone boson 
fields and $X^a$ the broken $SU(6)$ generators. The model assumes a gauged 
$SU(2)_1\times SU(2)_2$ subgroup of $SU(6)$ with generators 
($\sigma^a$ are the Pauli matrices) 
\beq 
Q_1^a = 
\left(\begin{array}{ccc} \sigma^a/2 & & \\ 
                           &  0_2 & \\
                           &  & 0_2  
 \end{array}\right)\ , 
\hspace{0.5cm} 
Q_2^a = 
\left(\begin{array}{ccc}  0_2 & & \\
                           &  -{\sigma^a_2}^*/2 & \\
                           &  & 0_2
\end{array}\right)\ .
\label{generatorsp} 
\eeq 
As usual, hypercharge can be embedded in different ways without 
destabilizing the 
Higgs mass due to the smallness of the $g'$ coupling. We focus therefore 
only in the $SU(2)$ part of the gauge sector. 
The vacuum expectation value in eq.~(\ref{vevsp}) breaks $SU(2)_1\times 
SU(2)_2$ down to the diagonal $SU(2)$, identified with the SM $SU(2)_L$ group.

The Goldstone and (pseudo)-Goldstone bosons in the hermitian matrix $\Pi$ 
in $\Sigma$ fall in representations of the SM group as 
\beq 
\Pi = 
{1\over \sqrt{2}}\left(\begin{array}{cccc} 
\xi\sqrt{2} & \varphi & H_2^{\dagger} & H_1^{\dagger}\\ 
\varphi^\dagger & \xi^*\sqrt{2} & -H_1^T & H_2^T\\ 
H_2 & -H_1^* & 0 & 0 \\
H_1 & H_2^* & 0 & 0
\end{array}\right)+{1\over \sqrt{12}}\zeta^0{\rm 
diag}(1,1,1,1,-2,-2)\ , 
\label{pisp} 
\eeq 
where $H_1=(h_1^-,h_1^0)$ and $H_2=(h_2^0,h_2^+)$ are Higgs 
doublets; $\varphi$ is an antisymmetric $2\times 2$ matrix
\be 
\varphi=\left[\begin{array}{cc} 0 & \phi^0 \\ 
-\phi^0 & 0  \end{array} 
\right]\ , 
\ee 
containing a singlet $\phi^0$. Then, $\zeta^0$ is a  
Goldstone that might be eaten to give mass to a heavy $U(1)$ field or 
remain in the physical spectrum, depending on the embedding of 
hypercharge, and $\xi$ is the real triplet of Goldstone bosons associated to 
$SU(2)_1\times SU(2)_2\rightarrow SU(2)$ breaking: 
\be 
\xi={1\over 2} \sigma^a\xi^a=
\left[ \begin{array}{cc} {1\over 2}\xi^0 & {1\over \sqrt{2}} \xi^+\\ 
{1\over \sqrt{2}} \xi^- & -{1\over 2}\xi^{0} \end{array} \right]\ . 
\ee 
All the fields in $\Pi$ as written above are canonically normalized.

The kinetic part of the Lagrangian is 
\beq 
{\cal L}_{kin} = 
\frac{f^2}{8} {\rm Tr}[(D_{\mu}\Sigma)(D^{\mu}\Sigma)^\dagger]\ , 
\label{Lkin} 
\eeq 
where \beq D_{\mu}\Sigma = \partial_{\mu} \Sigma - 
i\sum_{j=1}^2 g_j W_{j\mu}^a(Q_j^a \Sigma + \Sigma Q_j^{a T})\ ,
\eeq
with an additional $U(1)$ contribution that we neglect.

In this model, the additional fermions required to cancel the top 
quadratic divergences can be introduced in several ways \cite{SU6Sp6,GWS,HS}. 
We consider here the fermionic couplings chosen in \cite{HS}, which 
are quite similar to the $SU(5)/SO(5)$ choice described in the main text. Again
there is a coloured pair of new fermions $t',{t'}^c$ and the relevant part 
of the Lagrangian containing the top Yukawa coupling is given by 
\beq 
{\cal L}_{f} = {1\over 2}h_{1} f \epsilon_{ijk} \epsilon_{xy} 
\chi_i \Sigma_{jx} \Sigma_{ky} {u_3'}^{c} + h_{2} f t' {t'}^c + 
h.c., 
\label{Lfsp} 
\eeq 
where now $\chi_i = (t, b, 0, 0,0,t')$, the indices $i,j,k$ run 
through $\{1,2,6\}$  and $x,y$ from 3 to 4, and $\epsilon_{ijk}$ and 
$\epsilon_{xy}$ are the 
completely antisymmetric tensors of dimension 3 and 2, respectively.

As in the $SU(5)/SO(5)$ Little Higgs model, gauge and 
fermion loops induce operators in the scalar potential of the form, 
\bea 
{\cal O}_{V_i}(\Sigma) & = & f^4 g_i^2 
\sum_a {\rm Tr}[(Q_i^{a}\Sigma)(Q_i^{a}\Sigma)^*]  \ ,\\
{\cal O}_F(\Sigma) & = & - 3 f^4 h_1^2 
\epsilon^{wx}\epsilon_{yz}\Sigma_{iw}\Sigma_{jx} \Sigma^{iy *}\Sigma^{jz 
*}\ .
\eea
It is easy to show that 
\bea
\label{t2}
\sum_a {\rm Tr}[(Q_1^{a}\Sigma)(Q_1^{a}\Sigma)^*]&=&
{3\over 2}-{3\over 4}  {\rm Tr}[\Sigma_2^\dagger\Sigma_2]\ ,\\
\label{t1}
\sum_a {\rm Tr}[(Q_2^{a}\Sigma)(Q_2^{a}\Sigma)^*]&=&
{3\over 2}-{3\over 4}  {\rm Tr}[\Sigma_1^\dagger\Sigma_1]\ ,
\eea
where now 
\be
\Sigma_1=(\Sigma_{i'x'})\ ,\;\;\;\; 
\Sigma_2=(\Sigma_{i''x''})\ .
\ee
with $i'=\{1,2,5,6\}$ and $x'=\{3,4\}$;  $i''=\{3,4,5,6\}$, and
$x''=\{1,2\}$; and
\be
\label{t3}
\epsilon^{wx}\epsilon_{yz}\Sigma_{iw}\Sigma_{jx} \Sigma^{iy *}\Sigma^{jz
*}=2{\rm Det}[\Sigma_3^\dagger\Sigma_3]\ ,
\ee
with 
\be
\Sigma_3=(\Sigma_{ix})\ ,
\ee
and $i=\{1,2,6\}$, $x=\{3,4\}$.

Once these scalar operators are added to the Lagrangian we should
also consider the contribution of scalar loops to the quadratic divergence 
of the effective potential. As in the $SU(5)/SO(5)$ model, the 
fermion-induced operator ${\cal O}_F(\Sigma)$  generates in this way a 
scalar operator
\be
\label{t3p}
{\cal O}_S(\Sigma)\equiv {\rm Tr}[\Sigma_3^\dagger\Sigma_3]\ ,
\ee
which in this particular case was not induced already by gauge-boson loops. 

The reason behind this behaviour is of course that unlike what 
happened in the Littlest Higgs model, the symmetry 
properties of the fermion couplings in the Lagrangian are not those of the 
gauge boson sector. More explicitly, type-2 operators like (\ref{t2}) are 
invariant under a global $SU(4)_2$ symmetry and type-1 operators like 
(\ref{t1}) are invariant under a different global $SU(4)_1$ symmetry. The 
transformation properties of $\Sigma_{1,2}$ are
\be
\Sigma_1\rightarrow U_1 \Sigma_1 V_2^T\ ,\;\;\;
\Sigma_2\rightarrow U_2 \Sigma_1 V_1^T\ ,\;\;\;
\ee 
with $U_i$ a (global) $SU(4)_i$ matrix and $V_i$ a (local) $SU(2)_i$ 
matrix. The global $SU(4)_i$ symmetries guarantee the lightness of both 
Higgs doublets $H_1$ and $H_2$. The operators  (\ref{t3}) and  
(\ref{t3p}) are not of type-1 or type-2 but rather of a different type-3
with $\Sigma_3$ transforming as
\be
\Sigma_3\rightarrow U_3 \Sigma_1 V_2^T\ ,
\ee
with $U_3$ a matrix of a (global) $SU(3)_1$ which is a subgroup of 
$SU(4)_1$. This 
$SU(3)_1$ does not keep $H_2$ light and therefore, in the presence of 
type-3 operators, only $H_1$ will remain at the electroweak scale.

Writing the tree-level potential as
\be
V_0/f^4=
\kappa_1 {\mathrm Tr}[\Sigma_1^\dagger\Sigma_1]+
\kappa_2 {\mathrm Tr}[\Sigma_2^\dagger\Sigma_2]+
\kappa_3 {\mathrm Tr}[\Sigma_3^\dagger\Sigma_3]+
\kappa'_3 {\mathrm Det}[\Sigma_3^\dagger\Sigma_3]\ ,
\label{Vtree6}
\ee
we can compute the one-loop quadratically divergent contributions to this 
potential from loops of gauge bosons [neglecting again $U(1)$ corrections], 
fermions and scalars. We obtain
\bea
\delta V/f^4&=&-{\Lambda^2\over (4\pi f)^2}\left\{
\left({9\over 16}g_2^2+10\kappa_1\right) {\mathrm 
Tr}[\Sigma_1^\dagger\Sigma_1]+\right.
\left({9\over 16}g_1^2+10\kappa_2\right) {\mathrm
Tr}[\Sigma_2^\dagger\Sigma_2]\nonumber\\
&+&(10\kappa_3-8\kappa'_3) {\mathrm Tr}[\Sigma_3^\dagger\Sigma_3]
+4\kappa'_3 {\mathrm Tr}[\Sigma_3^\dagger\Sigma_3\Sigma_3^\dagger\Sigma_3]
+\left(6h_1^2+16\kappa'_3\right)
{\mathrm Det}[\Sigma_3^\dagger\Sigma_3]\nonumber\\
&+&\left. 4\kappa_3' \sigma_5\Sigma_3^\dagger\Sigma_3\sigma_5^\dagger
+4\kappa_3' d_5^\dagger\Sigma_3\Sigma_3^\dagger d_5{\over}\right\}\ ,
\label{main6}
\eea
where $d_5\equiv (\Sigma_{i5})$ and $\sigma_5\equiv(\Sigma_{5x})$ (with 
$i=\{1,2,6\}$ and $x=\{3,4\}$).  We see that new operators not present in 
(\ref{main6}) are generated and  one should also include them from the 
beginning for a complete analysis. 
For such complete analysis we should 
include the operators\footnote{We have used the relation ${\rm 
Det}[\Sigma_3^\dagger\Sigma_3]=(T^2_3- {\rm
Tr}[\Sigma_3^\dagger\Sigma_3\Sigma_3^\dagger\Sigma_3])/2$.}
$T_\alpha\equiv {\rm 
Tr}[\Sigma_\alpha^\dagger\Sigma_\alpha]$ ($\alpha=1,3$), $ {\rm
Tr}[\Sigma_3^\dagger\Sigma_3\Sigma_3^\dagger\Sigma_3]$, $T_3^2$, 
$d_5^\dagger\Sigma_3\Sigma_3^\dagger d_5$, 
$\sigma_5\Sigma_3^\dagger\Sigma_3 \sigma_5^\dagger$, $d_5^\dagger d_5$, 
$\sigma_5\sigma_5^\dagger$, $d_5^\dagger d_5 T_3$ and 
$\sigma_5\sigma_5^\dagger T_3$. Using the 
previous 
basis, the 
matrix $R$ 
of eq.~(\ref{R}) is 11-dimensional and breaks up in two blocks. The first 
$2\times 2$ block, in the space spanned by $\{T_1,T_2\}$, is simply 
$-10I_2$. The $9\times 9$ block is
\be
R=\left(\begin{array}{ccccccccc}
10 & º0 & 0 & 0 & 0 & 0 & 0 & 0 & 0 \\
-16& 16 & 4 & -4 & -4 & 0 & 0 & 0 & 0 \\
-32& 8 & 20 & 4& 4& 0 & 0 & 0 & 0 \\
-2 & 0 &    0 &20& 4& -4 & 0 & 4 & 0 \\
-2 & 0 &    0 &4 &20& 0 & -6 & 0  &4 \\
0 & 0 & 0 & 0 & 0 & 10 & 0 & 0 & 0 \\
0 & 0 & 0 & 0 & 0 & 0 & 10 & 0 & 0 \\
-6 & 0 &  0 &4 &-4& -12 & 0 & 20  & 0\\
-4 & 0 &  0 &-4  &4& 0& -12 &0 & 20  \\
\end{array}\right)\ ,
\ee
and has eigenvalues $\{10(3),24(3),12(2),20\}$ (with 
the multiplicities given in parenthesis). We can then deduce the bound
\be
\Lambda\simlt 4\pi f/\sqrt{24}\ .
\ee

Concerning vacuum alignment issues \cite{va0,va}, an expansion of the 
scalar potential
(\ref{Vtree6}) including the corrections in (\ref{main6}) gives
\bea
V&=&4f^2|\phi^0|^2\left\{{\over}\kappa_1+\kappa_2-(\kappa_3+\kappa_3')
\right.\nonumber\\
&+&\left.{\Lambda^2\over (4\pi f)^2}\left[-{9\over 16}(g_1^2+g_2^2)
+ 6 h_1^2-10(\kappa_1+\kappa_2)+10\kappa_3+16\kappa_3'
\right]\right\}\nonumber\\
&+&\left. 2f^2|H_2|^2\left\{-(\kappa_3+\kappa_3')+
{\Lambda^2\over (4\pi f)^2}\left[6h_1^2+10\kappa_3+12\kappa_3'
\right.\right]\right\}+... 
\eea
It is then easy to choose the unknown couplings $\kappa_\alpha$ and 
$\kappa_3'$ 
to overcome the negative correction from gauge boson loops (a known 
problem of this LH model) in such a way 
that both $\phi^0$ and $H_2$ have positive masses (of order $f$). 
Justifying such choice of couplings is only possible in the context of a 
UV completion of the model.

\section*{C. The ${\bma{SU(N)/SO(N)}}$ Case}
\setcounter{equation}{0} 
\renewcommand{\theequation}{C.\arabic{equation}}

In order to gain some understanding on how the large renormalization 
effects from scalar loops scale with the size of the groups involved, we 
consider here, as a simple exercise, the case of a Little Higgs model with 
coset structure $SU(N)/SO(N)$. We simply repeat the loop calculation in 
the mean text keeping track of the $N$-dependence.

We normalize the $SU(N)$ generators as
\be
{\rm Tr}(T^aT^b)=\delta_{ab}\ .
\ee
We then write the well known relations (see {\it e.g.}~\cite{cvitanovic})
\bea
\sum_{a \in SU(N)} T^a_{ij}T^a_{kl}&=&\delta_{il}\delta_{jk}-{1\over 
N}\delta_{ij}\delta_{kl}\ ,\nonumber\\
\sum_{a \in SO(N)} 
T^a_{ij}T^a_{kl}&=&{1\over 
2}(\delta_{il}\delta_{jk}-\delta_{ik}\delta_{jl})\ ,
\eea
from which we obtain
\be
\sum_{a \in SU(N)/SO(N)}T^a_{ij}T^a_{kl}\equiv
\sum_a [X_a^{(0)}]_{ij}[X_a^{(0)}]_{kl}={1\over 2}\delta_{il}\delta_{jk}+
{1\over 2}\delta_{ik}\delta_{jl}
-{1\over N}\delta_{ij}\delta_{kl}
\ ,\nonumber\\
\ee
where $a \in SU(N)/SO(N)$ runs over the broken generators, $X_a^{(0)}$.
The superindex $0$ refers to the original basis where the VEV producing 
the $SU(N)\rightarrow SO(N)$ breaking is simply the identity $I_5$. 
In that basis we then find (the diagrammatic techniques of 
\cite{cvitanovic} proved very useful in deriving these and similar 
identities) 
\be
\sum_a {\mathrm Tr}[X_a^{(0)} YX_a^{(0)} Z]=
\sum_a {\mathrm Tr}[X_a^{(0)} YX_a^{(0)*} Z]=
{1\over 2} {\mathrm Tr}[Y]{\mathrm Tr}[Z]
-{1\over N}{\mathrm Tr}[YZ]+{1\over 2}
{\mathrm Tr}[Y Z^T]\ .
\ee
In a rotated basis with generators $X_a=U_0X_a^{(0)}U_0^\dagger$
and VEV $\Sigma_0=U_0U_0^T$ we get
\bea
\sum_a {\mathrm Tr}[X_a YX_a^* Z]&=&
{1\over 2} {\mathrm Tr}[YZ^T]-{1\over N}{\mathrm Tr}[YZ]
+{1\over 2}{\mathrm Tr}[Y\Sigma_0^*]{\mathrm Tr}[Z\Sigma_0]\ 
,\nonumber\\
\sum_a {\mathrm Tr}[X_a YX_a Z]&=&
{1\over 2} {\mathrm Tr}[Y]{\mathrm Tr}[Z]
-{1\over N}{\mathrm Tr}[YZ]+{1\over 2}
{\mathrm Tr}[Y\Sigma_0Z^T \Sigma_0^*]\ ,
\label{identN}
\eea
of which eq.~(\ref{ident}) is a particular case with $N=5$.

Following the $SU(5)/SO(5)$ case we can gauge two $SU(2)\times U(1)$ 
subgroups of SU(N) so that, when $g_1=g'_1=0$ there is a global $SU(N-2)$ 
symmetry protecting the mass of the Higgs, etc. The submatrices $\Sigma_1$ 
and $\Sigma_2$ are generalized to $(N-2)\times 2$ matrices transforming as
$\Sigma_1\rightarrow U_1\Sigma_1 V_2^T$ and $\Sigma_2\rightarrow 
U_2\Sigma_2 V_1^T$, where $U_i$ is a matrix of the global $SU(N-2)_i$ and
$V_i$ a matrix of the local $SU(2)_i$. Writing down the tree-level 
scalar potential as in (\ref{Vtree}) we can again compute the contribution 
of divergent scalar loops to the same scalar operators obtaining
\bea
\delta_S\kappa_1&=&-\left[2(N+1)\kappa_1-2(N-3)\kappa'_1\right]{\Lambda^2\over
(4\pi f)^2}\ ,\nonumber\\
\delta_S\kappa_2&=&-\left[2(N+1)\kappa_2-2(N-3)\kappa'_2\right]{\Lambda^2\over 
(4\pi f)^2}\ ,\nonumber\\
\delta_S\kappa'_1&=&-4N\kappa'_1{\Lambda^2\over
(4\pi f)^2}\ ,\nonumber\\
\delta_S\kappa'_2&=&-4N\kappa'_2{\Lambda^2\over
(4\pi f)^2}\ .
\label{mainN}
\eea
From this result we conclude that the upper limit on the cut-off scale 
that follows from requiring consistency of the low-energy effective theory
approach is 
\be
\Lambda\simlt 4\pi f/\sqrt{4N}\ .
\ee
[The number of pseudo-Goldstones in this example is $(N^2+N-2)/2$.]

\section*{Acknowledgments}
We thank Ann Nelson for useful correspondence. This work is 
supported by the Spanish Ministry of Education and Science through a 
M.E.C. project (FPA2004-02015) and by a Comunidad de Madrid  
project (HEPHACOS; P-ESP-00346). Jos\'e Miguel No acknowledges the 
financial support of a FPU grant from M.E.C.

\end{document}